\documentclass[prd,twocolumn,showpacs,nofootinbib,preprintnumbers,floatfix]{revtex4}  
\usepackage[plainpages=false, colorlinks=true, anchorcolor=blue, linkcolor=blue, citecolor=blue, bookmarks=false]{hyperref}
\newcommand{\bqn}{\begin{eqnarray}}
\newcommand{\eqn}{\end{eqnarray}}
\usepackage{graphicx}  
\usepackage{dcolumn}   
\usepackage{bm}        
\usepackage{amssymb}   
\usepackage{natbib}
\usepackage{amsmath}   
\usepackage{longtable}  
\usepackage{float} \usepackage[utf8]{inputenc}
\usepackage{tasks}
\newcommand{\rthis}[1]{\textcolor{black}{#1}}

\begin{document}

\newcommand{\apjl}{Astrophys. J. Lett.}
\newcommand{\apjs}{Astrophys. J. Suppl. Ser.}
\newcommand{\aap}{Astron. \& Astrophys.}
\newcommand{\aj}{Astron. J.}
\newcommand{\pasp}{PASP}
\newcommand{\araa}{Ann. Rev. Astron. Astrophys. } 
\newcommand{\aapr}{Astronomy and Astrophysics Review}
\newcommand{\ssr}{Space Science Reviews}
\newcommand{\mnras}{Mon. Not. R. Astron. Soc.}
\newcommand{\apss} {Astrophys. and Space Science}
\newcommand{\jcap}{JCAP}
\newcommand{\na}{New Astronomy}
\newcommand{\pasj}{PASJ}
\newcommand{\pasa}{Pub. Astro. Soc. Aust.}
\newcommand{\physrep}{Physics Reports}

\title{A search for the  variation  of speed of light using galaxy cluster gas mass fraction measurements}

\author{I. E. C. R. Mendon\c{c}a$^{1}$}\email{ramalho.isaac@fisica.ufrn.br}

\author{Kamal Bora$^{2}$}\email{ph18resch11003@iith.ac.in}

\author{R. F. L. Holanda$^{1,3,4}$}\email{holandarfl@fisica.ufrn.br}

\author{Shantanu Desai$^{2}$}\email{shntn05@gmail.com}

\author{S. H. Pereira$^{5}$}\email{s.pereira@unesp.br}

\affiliation{$^1$Departamento de F\'{\i}sica, Universidade Federal do Rio Grande do Norte, Natal - Rio Grande do Norte, 59072-970, Brazil}

\affiliation{$^2$ Department of Physics, Indian Institute of Technology, Hyderabad, Kandi, Telangana-502284, India }

\affiliation{$^3$Departamento de F\'{\i}sica, Universidade Federal de Campina Grande, 58429-900, Campina Grande - PB, Brazil}

\affiliation{$^4$Departamento de F\'{\i}sica, Universidade Federal de Sergipe, 49100-000, Aracaju - SE, Brazil}

\affiliation{$^5$Departamento de F\'isica, Faculdade de Engenharia de Guaratinguet\'a, Universidade Estadual Paulista (UNESP), 12516-410, Guaratinguet\'a, SP, Brazil}

\begin{abstract}

In this paper, we implement a new   method to test the invariance of the speed of light ($c$)  as a function of redshift, by combining the measurements of galaxy cluster gas mass fraction, $H(z)$  from cosmic chronometers, and  Type-Ia supernovae (SNe Ia). In our analyses, we consider both  a  constant depletion factor (which corresponds to the ratio by which the cluster gas mass fraction is depleted with respect to the universal baryonic mean) and one  varying with redshift. We also consider  the influence of different $H_0$ estimates on our results. We look for a variation of  $c$, given by $c(z)=c_0(1+c_1z)$. We find a degeneracy between our final results on $c$ variation and the assumptions on the gas mass fraction depletion factor. Most of our analyses  indicate negligible variation of 
the speed of light.

\end{abstract}

\maketitle

\section{Introduction}

At the background level the $\Lambda$CDM model~\cite{Ratra08}, also known as  the Concordance model of cosmology,  is very successful in explaining the different phases of evolution of the universe, and agrees with most of the observational data~\cite{2020A&A...641A...6P}. However, there are still some unresolved issues in
the framework of such a cosmic concordance model,  for eg., small-scale problems \cite{Bull:2015stt,Peebles21}, the cosmic curvature
problem \cite{DiValentino:2019qzk}, the Hubble
tension \cite{DiValentino:2020zio,Lahav21,Julien,Verde}, inability to detect cold dark matter candidates in the laboratory~\cite{Merritt}, among others. { Some of these issues can be further scrutinized using new observational probes, such as the lensing  of the cosmic microwave background (CMB) \cite{Smith:2007rg}, addition of B-mode polarization \cite{SPTpol:2013omd},  and the kinetic Sunyaev-Zel’dovich effect \cite{Hand:2012ui}. To explain some of these deficiencies in the standard model, many alternative models have been proposed, such as adding new dynamic fluids modeled by different fields, modifications of Einstein's general relativity or even the inclusion of extra dimensions \cite{2008ARA&A..46..385F,Bull:2015stt,Weinberg2013,2016ARNPS..66...95J,Desai18}. }

Another specific class of alternative models consider the possibility of a variation of some of the fundamental constants of Physics, such as the fine-structure constant $\alpha$ \cite{King:2012id,Leefer:2013waa,vandeBruck:2015rma,Kotus:2016xxb,Goncalves:2019xtc,Galli:2012bf,Liu:2021mfk,Bora:2020sws,Colaco:2019fvl}, Newton's gravitational constant $G$ \cite{Dirac,Jofre:2006ug,Verbiest:2008gy,Lazaridis:2009kq,Garcia-Berro:2011kvq,Ooba:2016slp,Zhao:2018gwk,Vijaykumar:2020nzc} or the speed of light $c$ \cite{Zhu:2021pml,Liu:2018qrg,Xu:2016zsa,Xu:2016zxi,Cruz:2012bwp,FermiGBMLAT:2009nfe,Liu:2021eit,Cao:2018rzc,Cao:2016dgw,Rajdeep}. In this work, we focus on testing  the supposition of the  constancy of the speed of light, which constitutes one of the most basic and sacrosanct tenets of Physics. Any indication of a time varying speed of light has deep implications for both fundamental physics and cosmological models. The horizon and flatness problems in the standard big bang model, for instance can be solved  based on variations of $c$ \cite{Moffat93,Albrecht:1998ir}, providing an alternative to the Standard inflation scenario. Also, a possible method to explain the scale-invariant spectrum of CMB data using varying speed of light was demonstrated in  \cite{Magueijo:2003gj}.

On the other hand, it is important to draw attention to the fact that the motivation for a varying speed of light needs to be studied very carefully (see \cite{Ellis:2003pw} for a review). The nature of the speed of light is complex and can have different facets. Just as an example,
if it is the electromagnetic speed that is supposed to vary, one needs to show how Maxwell’s equations are to be
changed. Another possibility is that  \rthis{since the speed of light also plays the role of the limiting speed for any relative motion}, the causal relativistic speed could vary, \rthis{where the causal relativistic speed corresponds to a universal speed, which is invariant under velocity addition: $v + v_{lim} \rightarrow v_{lim}$~\cite{Ellis07}}.
\rthis{Therefore, if it is this aforementioned limiting speed which needs to be changed,} one should be able to show how the spacetime metric tensor changes. Some authors defend the idea that the time variation of fundamental constants is frequently presented in a misleading way and the time-variation in the physical laws
must best described in terms of time variation of dimensionless ratios, rather than 
constants with dimensions. More detailed discussions on these issues can be  found in   \cite{Ellis:2003pw,Duff:2001ba,Uzan:2002vq,Ellis07,Moffat08}. \rthis{In order to avoid such misconceived formulation, we consider our analysis in terms of the ratio $\Delta c/c$, which is independent of units.}

Despite the existence of serious conceptual problems faced while trying to change the status of a constant $c$ in Physics, several proposals to measure the constancy  of the speed of light on cosmic scales have been carried out recently. The main motivation is that  $c=299792458$ m/s value is only a local (at a redshift of zero) measure of the speed of light. Therefore, it is important to probe its variation with redshift.
Furthermore, the avalanche
of observational data in cosmology enables us to measure this variation in the distant universe with high precision. We briefly recap these results on the cosmological tests for the  constancy of speed of light.

Recently, a method to study the possible variation of $c$ by using Baryon Acoustic Oscillations
(BAO) was proposed by
 \cite{Salzano:2014lra}, based on a simple relation between the
angular diameter distance $D_A$ and the Hubble parameter
function $H(z)$ at the same redshift. In \cite{Cao:2016dgw},
the measurement of $c$ was done by using
$D_A$ from
intermediate-luminosity radio quasars calibrated as standard rulers, and the method was extended by \cite{Salzano:2016hce} to different redshifts.   In \cite{Qi:2014zja}, type Ia Supernova (SNeIa), BAO, H(z), and CMB data were combined,  and the  variation of speed of light was constrained to be of the order of $10^{-2}$. The same result was confirmed by \cite{Liu:2021eit} using a combination of strong gravitational lensing (SGL) systems and SNeIa. 
Ref.~\cite{Wang:2019tdn}  used a model-independent method to reconstruct the temporal evolution of the speed of light, and the results were in agreement to the value measured at $z=0$. Complementary to these tests on the  redshift-dependent speed of light, searches for an energy-dependent speed of light (as predicted by certain Lorentz-violating standard model extension models) have also been carried out using spectral lags of Gamma-ray bursts (See ~\cite{Rajdeep, Bartlett} and references therein.) 

In this work, we perform a consistency test for the  invariance of  speed of light with galaxy clusters, $H(z)$ measurements, and SNe Ia data. From galaxy cluster observations we use 40 X-ray gas mass fraction measurements in the redshift interval $0.16 \leq z \leq 1.016$ \cite{Mantz:2014xba}.  The model which we posit for the variation in the speed of light is given as: $c(z) \equiv c_0(1+c_1z) \equiv c_0\phi(z)$, where $c_0$ represents the speed of light at $z=0$.
We find that the  galaxy cluster gas mass fraction measurements are sensitive to a possible variation of the speed of light. 

The complementary probes we used for our analyses are: luminosity distances from $H(z)$ estimates~\cite{li19} and   SNe Ia data~\cite{pantheon}. We also explore the influence of different $H_0$ estimates on our results: Planck satellite~\cite{2020A&A...641A...6P} and local estimates \cite{2021ApJ...908L...6R}. All our analyses are compatible with no evolution for the depletion factor ($\gamma_1 \approx 0$ within 1$\sigma$ c.l.). In most of the scenarios explored hitherto, no variation of the speed of light was found ($c_1 = 0$ within 1$\sigma$ c.l.). When we consider $\gamma(z)=0.85 \pm 0.08$ plus luminosity distances  estimated by using Pantheon SNe Ia sample along with the $H_0$ prior from \cite{2021ApJ...908L...6R}, $c_1\neq 0$ even within 3$\sigma$ c.l. However, if the depletion factor is allowed to vary smoothly, $c_1=0$ is verified within 1$\sigma$ c.l.

The manuscript is organized as follows. The methodology adopted in this work is presented in Section~\ref{methodology}. In Section~\ref{data}, we briefly explain the cosmological data sample used in our analysis. Section~\ref{sec:analysis} describes our analysis and results. We conclude in Section~\ref{sec:conclusions}. 

\section{Methodology}
\label{methodology}
The baryonic matter content of galaxy clusters is dominated by the X-ray emitting intracluster gas, detected predominantly via  thermal bremmsstrahlung~\cite{sarazin}. A quantity of interest in cosmological analyses is the gas mass fraction, defined by $f_{gas}=M_{gas}/M_{tot}$, where $M_{gas}$ is the mass of the intracluster gas and $M_{tot}$ is the total mass, which includes baryonic gas mass and dark matter mass.

The gas mass $ M_{gas} (<R) $ within a radius $R$  obtained by X-ray observations can be written as \cite{sarazin}:
\begin{eqnarray}
M_{gas} (<R) &=& \left( \frac{3 \pi \hbar m_e c^2}{2 (1+X) e^6}
\right)^{1/2}  \left( \frac{3 m_e c^2}{2 \pi k_B T_e} \right)^{1/4}
m_H \nonumber\\
& & \mbox{\hspace{-2.5cm}} \times \frac{1}{[\overline{g_B}(T_e)]^{1/2}}
{r_c}^{3/2} \left
[ \frac{I_M (R/r_c, \beta)}{I_L^{1/2} (R/r_c, \beta)} \right] [L_X
(<R)]^{1/2}\;,
\end{eqnarray}
where $X$ is the hydrogen mass fraction, $T_e$ is the gas temperature, $m_e$ and $m_H$ are the electron and hydrogen masses, respectively, $\overline{g_B}(T_e)$ is the Gaunt factor,  $r_c$ stands for the core radius and
$$
I_M (y, \beta) \equiv \int_0^y (1+x^2)^{-3 \beta/2} x^2 dx\;,
$$
$$
I_L (y, \beta) \equiv \int_0^y (1+x^2)^{-3 \beta} x^2 dx\;.
$$
From Eq.(1),
\begin{equation}
M_{gas} (<\theta) \propto c^{3/2}.
\end{equation}
On the other hand, the total mass  within a given radius $R$ can be obtained by assuming that the intracluster gas is in hydrostatic equilibrium, i.e.~\cite{Allen2011}
\begin{equation}
M_{tot} (<R) = - \left. \frac{k_B T_e R}{G \mu m_H} \frac{d \ln
n_e(r)}{d \ln r} \right|_{r=R}.
\end{equation}
Therefore, from Eq.(1),  if $c(z) \equiv c_0\phi(z)$, the gas mass fraction defined earlier  in its more general form is given by:
\begin{equation}
\label{final}
f_{gas} \equiv \frac{\phi(z)^{3/2}M_{gas}}{M_{tot}}.
\end{equation}

Usually, the expected constancy of the $f_{gas}$ within massive, hot and relaxed galaxy clusters  can be used to constrain cosmological parameters by using the following equation (see, for instance, \cite{Allen2007,Ettori2009,Allen2011,Mantz:2014xba,Holanda2020}):

\begin{equation}
\label{fgas1}
f_{gas}(z) = \gamma(z)K(z) A(z) \left[\frac{\Omega_b}{\Omega_M}\right] \left(\frac{D_L^*}{D_L}\right)^{3/2}\,.
\end{equation}
Here, the asterisk denotes the corresponding quantities for the fiducial model used in the observations to obtain  $f_{gas}$ (usually a flat $\Lambda$CDM model with Hubble constant $H_0=70$ km s$^{-1}$ Mpc$^{-1}$ and the present-day total matter density parameter $\Omega_M=0.3$), $\gamma(z)$ is the depletion parameter, which indicates the amount of cosmic baryons that are thermalized within the cluster potential (see details in the Refs. \cite{Allen2007,Allen2011,Mantz:2014xba,2013ApJ...777..123B,2013MNRAS.431.1487P}). $K(z)$ stands for the calibration constant which is equal to $0.96\pm0.12$~\cite{Mantz:2014xba} and $A(z)$ represents the angular correction factor which is close to unity. The cosmological analyses with gas mass fraction measurements are model-independent due to the ratio in the parenthesis of Eq.~(\ref{fgas1}), which takes into account the expected variation in the gas mass fraction measurement when the underlying cosmology is varied.  \rthis{We do our analysis using two different assumptions on the gas depletion factor, involving  a redshift-dependent depletion factor: $\gamma(z)=\gamma_0(1+\gamma_1z)$,  as well as  a constant factor, whose value was obtained from hydrodynamical simulations ($\gamma(z)=0.85 \pm 0.08$) \cite{2013ApJ...777..123B,2013MNRAS.431.1487P}}. 

\rthis{Although simulations suggest a constant depletion factor~\cite{2013ApJ...777..123B,2013MNRAS.431.1487P},  in recent years cosmology agnostic measurements of $\gamma$ using gas mass fraction measurements  at $R_{500}$~\cite{BoraDesai21,Zheng19} and $R_{2500}$~\cite{2017JCAP...12..016H,Holanda21}  have been carried out,  
which hint towards  redshift-dependent $\gamma(z)$. Recently, Ref.~\cite{BoraDesai21} showed  that $\gamma (z)$ at $R_{500}$ is sample-dependent and is not consistent between Planck and SPT, but varies with redshift for both the samples.
For $R_{2500}$, Ref.~\cite{2017JCAP...12..016H} found no evidence for a varying $\gamma(z)$ 
when using distance measurements from Type 1a supernova. However, 
recently~\cite{Holanda21} found $2.7\sigma$ evidence for $\gamma (z)$ (using $R_{2500}$ measurements) decreasing with redshift, when the distance corresponding to the galaxy cluster redshift was obtained using strong lensing systems.}

For a redshift-dependent speed of light give by $c(z)=c_0\phi(z)$, this equation  would have to be modified accordingly to
\begin{equation}
\label{fgas2}
f_{gas}(z) = \phi(z)^{3/2}\gamma(z)K(z) A(z) \left[\frac{\Omega_b}{\Omega_M}\right] \left(\frac{D_L^*}{D_L}\right)^{3/2}\,.
\end{equation}
Then, if one knows the luminosity distance to a  galaxy cluster, it is possible to obtain limits on $\phi(z)$. In our method, the luminosity distance for each galaxy cluster of the sample  is obtained by using $H(z)$ cosmic chronometer measurements and SNe Ia as discussed below. In our analyses, we consider $\phi(z)=(1+c_1z)$.

\section{Cosmological data}
\label{data}
 
 \subsection{Gas Mass Fraction}

The Chandra X-ray sample used for this analysis consists of 40 galaxy clusters~\cite{Mantz:2014xba},  identified through a comprehensive search of the Chandra archive for hot ($kT \geq  5$ keV), massive and morphologically relaxed systems. The galaxy clusters span the redshift range $0.078 \leq z \leq 1.063$ (see the left panel of Fig.~\ref{fig:gmf}). 
The choice of relaxed systems minimize the systematic biases in the hydrostatic masses. In this sample the innermost regions of the clusters are excluded, and the gas mass fraction $f_{gas}$ is calculated in the \rthis{spherical shell $0.8 \leq r/r_{2500} \leq 1.2$.} A more detailed discussion about the data can be found in~\cite{Mantz:2014xba}. The value for $\Omega_b/\Omega_M$ used in Eq.~\ref{fgas2} was obtained  from  Planck 2020 Cosmology results \cite{2020A&A...641A...6P}. On the other hand,  most  cosmological constraints from X-ray emitting gas mass fraction observations have relied on hydrodynamical simulations \cite{2013ApJ...777..123B,2013MNRAS.431.1487P}, which have been used to link the
observed  gas mass fraction to the cosmic baryon fraction through the so-called depletion factor, i.e., $ \gamma= f_{gas}(\Omega_b/\Omega_M)^{-1}$, which in principle may be a function of redshift. \footnote{The first self-consistent observational constraint on the gas depletion factor at $r_{2500}$ was found by  Ref.~\cite{2017JCAP...12..016H}, combining  X-ray emitting gas mass fraction measurements and luminosity distance measurements from type Ia supernovae. As basic result, a constant depletion factor was found with value in full agreement with simulations.}. Therefore, in order to be conservative, we assume two possibilities for the depletion factor:  a constant value: $\gamma(z)=0.85 \pm 0.08$ \cite{2013ApJ...777..123B,2013MNRAS.431.1487P}, and an evolving one, such as $\gamma(z)=\gamma_0(1+\gamma_1z)$, similar to some of our previous works~\cite{Holanda21,Mendonca21}.

\subsection{Cosmic Chronometers}

We use 31 cosmic chronometer $H(z)$ data from Ref.~\cite{li19} in the redshift range $0.07 \leqslant z \leqslant 1.965 $ in order to derive the luminosity distance for each galaxy cluster.  Briefly, the age difference between passively evolving galaxies at different redshifts are calculated in order to obtain  the Hubble parameter $H(z)$ \cite{jimenez}. Cosmic chronometers (CC) are one of the most widely used model-agnostic probes  for deducing the observational value of the Hubble parameter at different redshifts (See ~\cite{Haveesh,Vagnozzi} and references therein for more details).

In order to derive the luminosity distance to each galaxy cluster, we choose   Gaussian Processes Regression~\cite{Seikel}, which is a non-parametric technique used to reconstruct a function at any input value, based on a set of measurements.  More details on the Gaussian Process based interpolation can be found in ~\cite{Haveesh,BoraDesai21,BoraDesaiCDDR}, and we followed the same implementation as in these aforementioned works.

The reconstructed luminosity distance is obtained via,
\begin{equation}
D_{L} (z) =  c\left(1+z\right)\int_{0}^{z}\frac{dz^{'}}{H(z')},
\label{eq:daz}
\end{equation}
where $H(z')$ is the non-parametric reconstruction of the Hubble parameter using Gaussian Processes regression. This reconstruction of $D_L$ using the $H(z)$ measurements can be found in the  right panel of Fig.~\ref{fig:gmf}.

 \subsection{SNe Ia sample}
 
 The SNe Ia sample used in our analyses is the so called Pantheon \cite{pantheon} sample, consisting of 1049 spectroscopically confirmed SNe Ia covering a redshift range of $0.01 \leq z \leq 2.3$. This dataset is the latest  state of the art sample of SNe Ia measurements  available in  literature.
 
\begin{figure*}
    \centering
    \includegraphics[width=8.3cm, height=6.5cm]{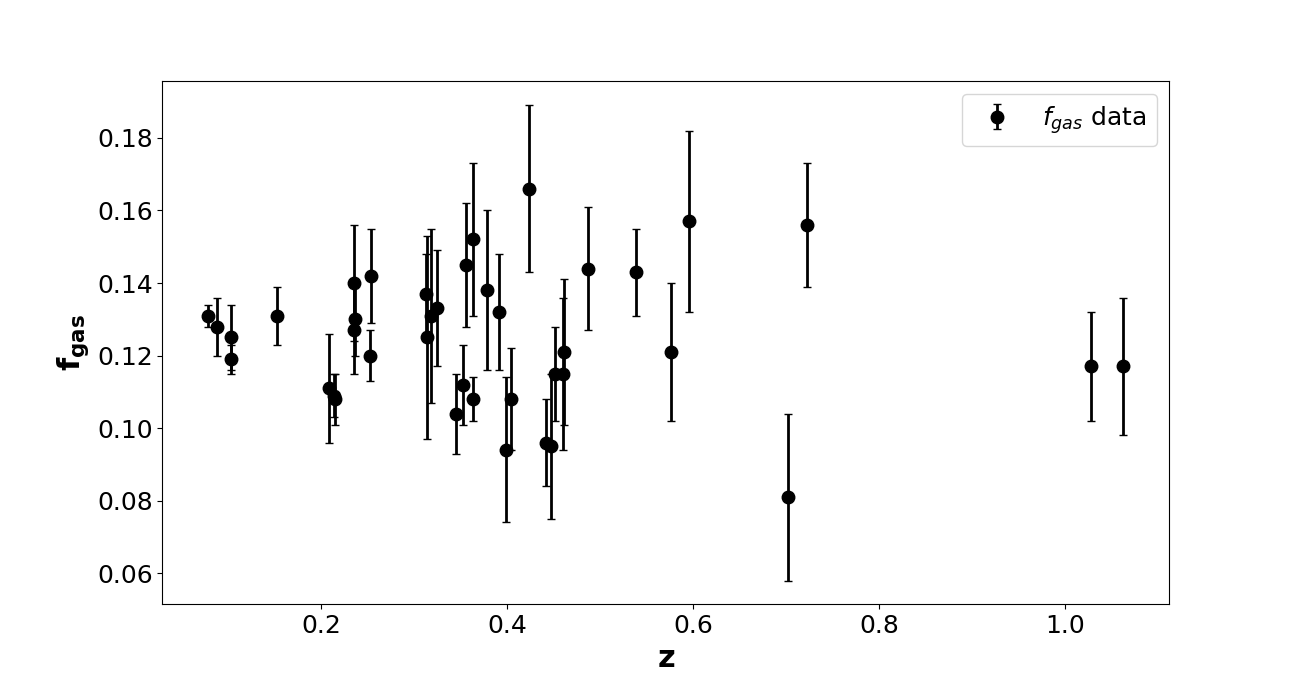} 
    \includegraphics[width=8.3cm, height=6.5cm]{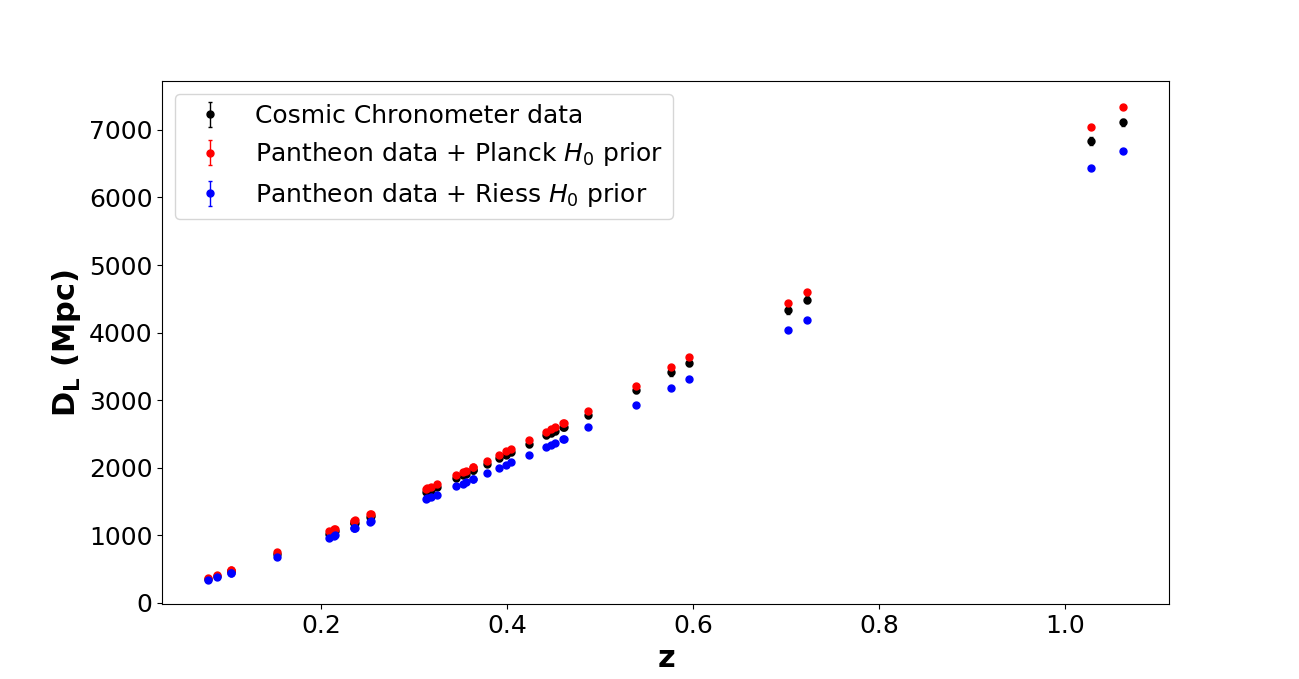} 
    \caption{(Left)The 40 Chandra X-ray gas mass fraction as a function of redshift compiled by \citet{Mantz:2014xba}. (Right) This figure displays the luminosity distance as a function of redshift estimated by different distance proxies as mentioned in the legend.  }
   \label{fig:gmf}
\end{figure*}

 However, to perform the appropriate tests, we must use SNe Ia at the same  redshifts as that of  the galaxy clusters. For this purpose we apply the Gaussian process method to find out the central value with the corresponding variance. Actually, we reconstruct two sets of $D_L$ measurements from the Pantheon sample~\cite{pantheon}. For the first sample, we consider the absolute magnitude $M_b = -19.23 \pm 0.04$   via the relation:
\begin{equation}
D_L=10^{(m_b-M_b-25)/5} \text{ Mpc}.
\end{equation}
This value is obtained by  assuming the Hubble constant provided by the Cepheids/SNe Ia estimates, $H_0 = 74.03 \pm 1.42$ km/s/Mpc  \cite{2021ApJ...908L...6R} (henceforth known as Riess prior).  The second one, is constructed by using $M_b = -19.43 \pm 0.02$, which is obtained using the Hubble constant estimated by the Planck Collaborations in the context of a $\Lambda$CDM model, $H_0 = 67.36 \pm 0.54$ km/s/Mpc \cite{2020A&A...641A...6P} (henceforth known as the Planck prior).  All the three distance estimates can be found in  the right panel of Fig.~\ref{fig:gmf}.

\vspace{0.2cm}

\section{Analysis and Results} 
\label{sec:analysis}
By using  Eq.(6), the constraints on the model parameters ($c_1$ or/and $\gamma_1$) can be obtained by maximizing the likelihood function, ${\cal{L}}$  given by

\begin{widetext}
\begin{equation}
    \label{eq:logL2}
   -2\ln\mathcal{L} = \sum_{i=1}^{n} \ln 2\pi{\sigma_{i}^2}+ \sum_{i=1}^{n}\frac{\left(\zeta(z)- \left[\frac{f_{gas}}{K(z) A(z) \gamma_0}\right]\left[\frac{\Omega_m}{\Omega_b}\right]\left[\frac{D_L}{D^*_L}\right]^{3/2}\right)^2}{\sigma_{i}^2} .
\end{equation}          \end{widetext} 

Here, $\sigma_i$  denotes the statistical errors associated with the gas mass fraction measurements and Pantheon sample/Cosmic Chronometers data, and is obtained by using standard error propagation methods.  
We choose the  $\tt{emcee}$ MCMC sampler~\cite{emcee} to maximize the log-likelihood function.  

We consider two cases for $\zeta(z)$:
\begin{itemize}
\item $\zeta(z) =   (1+c_1 z)^{3/2}$, along with a constant gas depletion factor, $\gamma (z) = 0.85\pm 0.08$ \cite{2013ApJ...777..123B,2013MNRAS.431.1487P}. 
\item   $\zeta(z) =  (1+\gamma_1 z) (1+c_1 z)^{3/2}$,  where the gas depletion factor varies according to the parametric form: $\gamma(z) = \gamma_0(1+\gamma_1 z)$, where $\gamma_0 =  0.85\pm0.08$ and $\gamma_1$ encapsulates a possible redshift evolution of the gas depletion factor, and  is treated as a free parameter in this case.
\end{itemize}
For each of the above two cases, we calculate the luminosity distance using all the three methods enumerated.  A summary can be found in Table~\ref{tab:table1}. We now summarize our results:

\begin{figure}
    \centering
    \includegraphics[width=7cm, height=7cm]{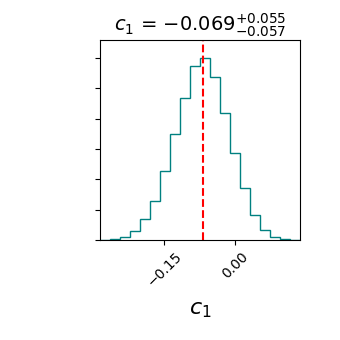} 
    \caption{ The likelihood distribution of parameter $c_1$ obtained using the {\tt Corner} python module~\cite{corner} by considering a constant depletion factor. The luminosity distance $D_L$ was estimated by using cosmic chronometer data.}
   \label{fig:cc1}
   \end{figure}

\begin{figure}
    \centering
    \includegraphics[width=7cm, height=7cm]{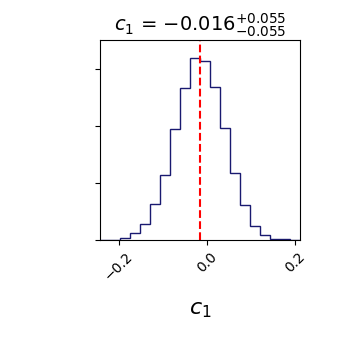} 
    \caption{ The likelihood distribution of parameter $c_1$ obtained using the {\tt Corner} python module~\cite{corner} by considering a constant depletion factor. The luminosity distance $D_L$ was estimated by using Pantheon sample along with the Planck $H_0$ prior \cite{2020A&A...641A...6P}.}
    \label{fig:planck1}
\end{figure}

\begin{figure}
    \centering
    \includegraphics[width=7cm, height=7cm]{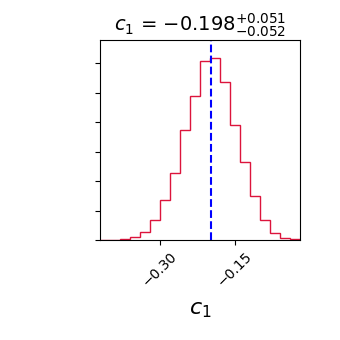} 
    \caption{ The likelihood distribution of parameter $c_1$ obtained using the {\tt Corner} python module~\cite{corner} by considering a constant depletion factor. The luminosity distance $D_L$ was estimated by using the Pantheon sample along with Riess $H_0$ prior \cite{2021ApJ...908L...6R}.}
    \label{fig:riess1}
\end{figure}

\begin{figure}
    \centering
    \includegraphics[width=9cm, height=8cm]{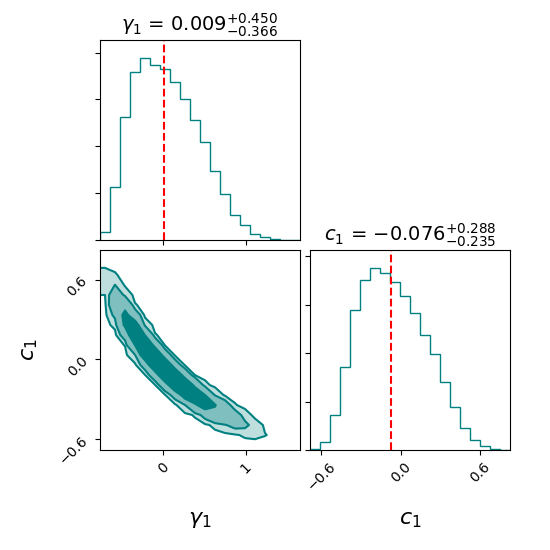} 
    \caption{The 1D marginalized posterior distributions along with 2D marginalized confidence intervals, obtained using the {\tt Corner} python module~\cite{corner}. The luminosity distance $D_L$ was estimated from cosmic chronometers.}
    \label{fig:cc2}
\end{figure}

\begin{figure}
    \centering
    \includegraphics[width=9cm, height=8cm]{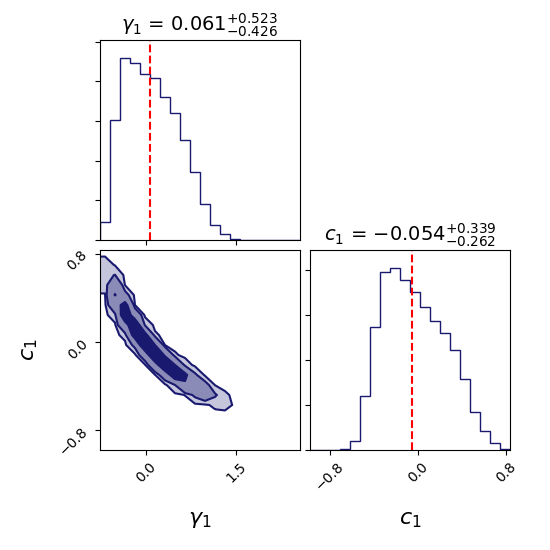} 
    \caption{The 1D marginalized posterior distributions along with 2D marginalized confidence intervals, obtained using the {\tt Corner} python module~\cite{corner}. The luminosity distance $D_L$ was estimated by using Pantheon sample along with the Planck $H_0$ prior \cite{2020A&A...641A...6P}.}
    \label{fig:planck2}
\end{figure}

\begin{figure}
    \centering
    \includegraphics[width=9cm, height=8cm]{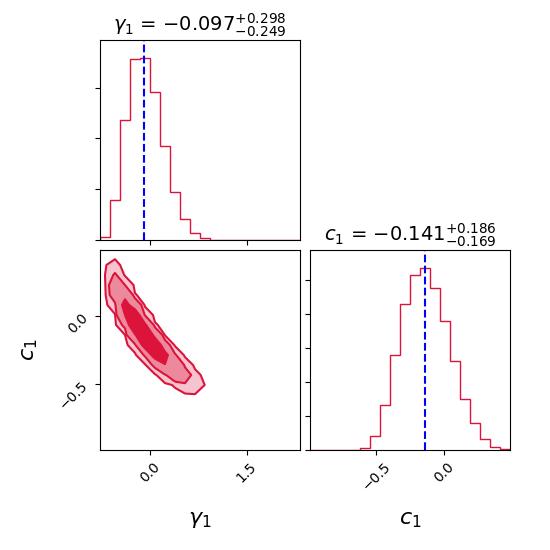} 
    \caption{The 1-D marginalized posterior distributions along with 2-D marginalized confidence intervals, obtained using the {\tt Corner} python module~\cite{corner}. The luminosity distance $D_L$ was estimated by using Pantheon sample along with the Riess $H_0$ prior \cite{2021ApJ...908L...6R}.}
    \label{fig:riess2}
\end{figure}

\begin{table*}[]
\caption{\label{tab:table1} Constraints on the parameter  $c_1$. }
    \centering
    \begin{tabular}{|l|c|c|c|c|r|} \hline
    \textbf{Distance Indicator} & $\gamma(z)$ considered  & \boldmath$\gamma_1$& \boldmath$c_1$\\ \hline 
      Cosmic Chronometers & $0.85\pm0.08$ & - &$-0.069\pm0.056$ \\
      Cosmic Chronometers & $0.85(1+\gamma_1z)$   & $0.009^{+0.45}_{-0.366}$& $-0.076\pm0.262$ \\
      Pantheon Sample with Planck $H_0$ prior & $0.85\pm0.08$ &-&  $-0.016\pm0.055$ \\
      Pantheon Sample with Planck $H_0$ prior & $0.85(1+\gamma_1z)$ &  $0.061^{+0.523}_{-0.426}$&$-0.054\pm0.301$ \\
      Pantheon Sample with Riess $H_0$ prior & $0.85\pm0.08$  &   - &$-0.198\pm0.052$\\
      Pantheon Sample with Riess $H_0$ prior & $0.85(1+\gamma_1z) $& $-0.097^{+0.298}_{-0.249}$& $-0.141\pm0.178$\\

      \hline 
      
    \end{tabular}

\end{table*}

\begin{itemize}
    \item {\bf Results for constant gas depletion factor:}
Fig.~\ref{fig:cc1}, Fig.~\ref{fig:planck1}, and Fig.~\ref{fig:riess1} show the 1-D likelihood distribution of the parameter $c_1$. In each case, the luminosity distance $D_L$ is estimated by using CC data, Pantheon data with the Planck $H_0$ prior, and the Riess $H_0$ prior, respectively. We find that $c_1$ estimated using the Pantheon sample with the Planck prior is consistent with no variation with redshift, to within 1$\sigma$. However, the other two cases show a mild decreasing trend for $c_1$ with redshift  at 1.2$\sigma$ using cosmic chronometer data, and at 3.9$\sigma$ using the Pantheon sample with Riess prior.

\item {\bf Results with an evolving gas depletion parameter:}  Our results after assuming an evolving gas depletion factor
can be found in  Fig.~\ref{fig:cc2}, Fig.~\ref{fig:planck2}, and Fig.~\ref{fig:riess2}, which show the 68\%, 95\%, and 99\% 2-D marginalized confidence intervals for $c_1$ and $\gamma_1$.
These correspond to the luminosity distance $D_L$ been estimated by using cosmic chronometer data, Pantheon data with Planck $H_0$ prior, and Riess $H_0$ prior, respectively. We find that $c_1$ is consistent with 0 within 1$\sigma$ for all these three cases considered. Therefore, we find that there is no redshift-dependent variation in the speed of light, once we consider an evolving gas depletion factor.
The sensitivity we obtain on $\Delta c/ c$  is $\Delta c/ c \approx \mathcal{O} (10^{-2})$,   in agreement with the independent recent estimates  \cite{Qi:2014zja,Wang:2019tdn,Liu:2021eit}.
\end{itemize}

\section{Conclusions}
\label{sec:conclusions}

In recent years, the increased availability of high quality observational data has allowed us  to self-consistently test  some of the fundamental tenets of the standard model of Cosmology. The invariance of the speed of light ($c$) with redshift is the question we have investigated in this work. As commented earlier, any indication of a time varying speed of light has deep implications for both fundamental Physics and cosmological models beyond $\Lambda$CDM.

In this paper, the invariance of the speed of light was tested by combining galaxy cluster gas mass fraction measurements, $H(z)$ measurements from cosmic chronometers, and SNe Ia observations from the Pantheon sample.  We have considered two {\it ansatz} for the  depletion factor ($\gamma(z)$):  a constant value, as well as  evolving with redshift according to  $\gamma(z)=\gamma_0(1+\gamma_1z)$. We search for a redshift-dependent   $c$, given by $c(z)=c_0(1+c_1z)$.  We also use three different measures of distance: one obtained using cosmic chronometers, and two from the Pantheon sample, corresponding to two different $H_0$ priors. 

Our results using these three distance measures for a constant gas fraction can be found in Figs.~\ref{fig:cc1}, ~\ref{fig:planck1}, and ~\ref{fig:riess1}. We find that the speed of light decreases with  redshift  at 1.2$\sigma$ using chronometer data, and at 3.9$\sigma$ using the Pantheon sample with Riess prior. It is however consistent with no variation, when we use the Pantheon sample coupled with the Planck prior. Our results for the  redshift-dependent gas depletion factors can be found in Figs.~\ref{fig:cc2},~\ref{fig:planck2}, and ~\ref{fig:riess2}. In this case, we find that our results are consistent with no variation of $c$ to within 1$\sigma$. A summary of all our results are compiled in Table~\ref{tab:table1}. Therefore, we find that there is a degeneracy between our results on the variation of $c$ and the assumptions used for the gas depletion factor. The sensitivity to variation of speed of light is given by $\Delta c/ c \approx \mathcal{O} (10^{-2})$, in full agreement with recent independent estimates \cite{Qi:2014zja,Wang:2019tdn,Liu:2021eit}.

Finally, we stress that the combinations of  galaxy cluster gas mass fraction plus $H(z)$ measurements and galaxy cluster gas mass fraction in conjunction with  SNe Ia
provide  a novel way to perform a consistency test for the invariance of speed of light. Furthermore, the results from  galaxy cluster gas mass fraction plus $H(z)$ are  independent of any calibrators usually adopted in the determinations of the
distance scale.

\section*{ACKNOWLEDGEMENT}
KB acknowledges the Department of Science and Technology, Government of India for providing the financial support under DST-INSPIRE Fellowship program. RFLH
thanks CNPq No.428755/2018-6 and 305930/2017-6. SHP acknowledges financial support from  {Conselho Nacional de Desenvolvimento Cient\'ifico e Tecnol\'ogico} (CNPq)  (No. 303583/2018-5). \rthis{We also thank the anonymous referee for useful constructive feedback on the manuscript.}

\bibliography{ref}

\begin{thebibliography}{80}
\expandafter\ifx\csname natexlab\endcsname\relax\def\natexlab#1{#1}\fi
\expandafter\ifx\csname bibnamefont\endcsname\relax
  \def\bibnamefont#1{#1}\fi
\expandafter\ifx\csname bibfnamefont\endcsname\relax
  \def\bibfnamefont#1{#1}\fi
\expandafter\ifx\csname citenamefont\endcsname\relax
  \def\citenamefont#1{#1}\fi
\expandafter\ifx\csname url\endcsname\relax
  \def\url#1{\texttt{#1}}\fi
\expandafter\ifx\csname urlprefix\endcsname\relax\def\urlprefix{URL }\fi
\providecommand{\bibinfo}[2]{#2}
\providecommand{\eprint}[2][]{\url{#2}}

\bibitem[{\citenamefont{{Ratra} and {Vogeley}}(2008)}]{Ratra08}
\bibinfo{author}{\bibfnamefont{B.}~\bibnamefont{{Ratra}}} \bibnamefont{and}
  \bibinfo{author}{\bibfnamefont{M.~S.} \bibnamefont{{Vogeley}}},
  \bibinfo{journal}{\pasp} \textbf{\bibinfo{volume}{120}}, \bibinfo{pages}{235}
  (\bibinfo{year}{2008}), \eprint{0706.1565}.

\bibitem[{\citenamefont{{Planck Collaboration}
  et~al.}(2020)\citenamefont{{Planck Collaboration}, {Aghanim}, {Akrami},
  {Ashdown}, {Aumont}, {Baccigalupi}, {Ballardini}, {Banday}, {Barreiro},
  {Bartolo} et~al.}}]{2020A&A...641A...6P}
\bibinfo{author}{\bibnamefont{{Planck Collaboration}}},
  \bibinfo{author}{\bibfnamefont{N.}~\bibnamefont{{Aghanim}}},
  \bibinfo{author}{\bibfnamefont{Y.}~\bibnamefont{{Akrami}}},
  \bibinfo{author}{\bibfnamefont{M.}~\bibnamefont{{Ashdown}}},
  \bibinfo{author}{\bibfnamefont{J.}~\bibnamefont{{Aumont}}},
  \bibinfo{author}{\bibfnamefont{C.}~\bibnamefont{{Baccigalupi}}},
  \bibinfo{author}{\bibfnamefont{M.}~\bibnamefont{{Ballardini}}},
  \bibinfo{author}{\bibfnamefont{A.~J.} \bibnamefont{{Banday}}},
  \bibinfo{author}{\bibfnamefont{R.~B.} \bibnamefont{{Barreiro}}},
  \bibinfo{author}{\bibfnamefont{N.}~\bibnamefont{{Bartolo}}},
  \bibnamefont{et~al.}, \bibinfo{journal}{\aap} \textbf{\bibinfo{volume}{641}},
  \bibinfo{eid}{A6} (\bibinfo{year}{2020}), \eprint{1807.06209}.

\bibitem[{\citenamefont{Bull et~al.}(2016)}]{Bull:2015stt}
\bibinfo{author}{\bibfnamefont{P.}~\bibnamefont{Bull}} \bibnamefont{et~al.},
  \bibinfo{journal}{Phys. Dark Univ.} \textbf{\bibinfo{volume}{12}},
  \bibinfo{pages}{56} (\bibinfo{year}{2016}), \eprint{1512.05356}.

\bibitem[{\citenamefont{{Peebles}}(2021)}]{Peebles21}
\bibinfo{author}{\bibfnamefont{P.~J.~E.} \bibnamefont{{Peebles}}},
  \bibinfo{journal}{arXiv e-prints} \bibinfo{eid}{arXiv:2106.02672}
  (\bibinfo{year}{2021}), \eprint{2106.02672}.

\bibitem[{\citenamefont{Di~Valentino et~al.}(2019)\citenamefont{Di~Valentino,
  Melchiorri, and Silk}}]{DiValentino:2019qzk}
\bibinfo{author}{\bibfnamefont{E.}~\bibnamefont{Di~Valentino}},
  \bibinfo{author}{\bibfnamefont{A.}~\bibnamefont{Melchiorri}},
  \bibnamefont{and} \bibinfo{author}{\bibfnamefont{J.}~\bibnamefont{Silk}},
  \bibinfo{journal}{Nature Astron.} \textbf{\bibinfo{volume}{4}},
  \bibinfo{pages}{196} (\bibinfo{year}{2019}), \eprint{1911.02087}.

\bibitem[{\citenamefont{Di~Valentino et~al.}(2021)}]{DiValentino:2020zio}
\bibinfo{author}{\bibfnamefont{E.}~\bibnamefont{Di~Valentino}}
  \bibnamefont{et~al.}, \bibinfo{journal}{Astropart. Phys.}
  \textbf{\bibinfo{volume}{131}}, \bibinfo{pages}{102605}
  (\bibinfo{year}{2021}), \eprint{2008.11284}.

\bibitem[{\citenamefont{{Shah} et~al.}(2021)\citenamefont{{Shah}, {Lemos}, and
  {Lahav}}}]{Lahav21}
\bibinfo{author}{\bibfnamefont{P.}~\bibnamefont{{Shah}}},
  \bibinfo{author}{\bibfnamefont{P.}~\bibnamefont{{Lemos}}}, \bibnamefont{and}
  \bibinfo{author}{\bibfnamefont{O.}~\bibnamefont{{Lahav}}},
  \bibinfo{journal}{arXiv e-prints} \bibinfo{eid}{arXiv:2109.01161}
  (\bibinfo{year}{2021}), \eprint{2109.01161}.

\bibitem[{\citenamefont{{Sch{\"o}neberg}
  et~al.}(2021)\citenamefont{{Sch{\"o}neberg}, {Abell{\'a}n}, {P{\'e}rez
  S{\'a}nchez}, {Witte}, {Poulin}, and {Lesgourgues}}}]{Julien}
\bibinfo{author}{\bibfnamefont{N.}~\bibnamefont{{Sch{\"o}neberg}}},
  \bibinfo{author}{\bibfnamefont{G.~F.} \bibnamefont{{Abell{\'a}n}}},
  \bibinfo{author}{\bibfnamefont{A.}~\bibnamefont{{P{\'e}rez S{\'a}nchez}}},
  \bibinfo{author}{\bibfnamefont{S.~J.} \bibnamefont{{Witte}}},
  \bibinfo{author}{\bibfnamefont{c.~V.} \bibnamefont{{Poulin}}},
  \bibnamefont{and}
  \bibinfo{author}{\bibfnamefont{J.}~\bibnamefont{{Lesgourgues}}},
  \bibinfo{journal}{arXiv e-prints} \bibinfo{eid}{arXiv:2107.10291}
  (\bibinfo{year}{2021}), \eprint{2107.10291}.

\bibitem[{\citenamefont{{Verde} et~al.}(2019)\citenamefont{{Verde}, {Treu}, and
  {Riess}}}]{Verde}
\bibinfo{author}{\bibfnamefont{L.}~\bibnamefont{{Verde}}},
  \bibinfo{author}{\bibfnamefont{T.}~\bibnamefont{{Treu}}}, \bibnamefont{and}
  \bibinfo{author}{\bibfnamefont{A.~G.} \bibnamefont{{Riess}}},
  \bibinfo{journal}{Nature Astronomy} \textbf{\bibinfo{volume}{3}},
  \bibinfo{pages}{891} (\bibinfo{year}{2019}), \eprint{1907.10625}.

\bibitem[{\citenamefont{Merritt}(2017)}]{Merritt}
\bibinfo{author}{\bibfnamefont{D.}~\bibnamefont{Merritt}},
  \bibinfo{journal}{Stud. Hist. Phil. Sci. B} \textbf{\bibinfo{volume}{57}},
  \bibinfo{pages}{41} (\bibinfo{year}{2017}), \eprint{1703.02389}.

\bibitem[{\citenamefont{Smith et~al.}(2007)\citenamefont{Smith, Zahn, and
  Dore}}]{Smith:2007rg}
\bibinfo{author}{\bibfnamefont{K.~M.} \bibnamefont{Smith}},
  \bibinfo{author}{\bibfnamefont{O.}~\bibnamefont{Zahn}}, \bibnamefont{and}
  \bibinfo{author}{\bibfnamefont{O.}~\bibnamefont{Dore}},
  \bibinfo{journal}{Phys. Rev. D} \textbf{\bibinfo{volume}{76}},
  \bibinfo{pages}{043510} (\bibinfo{year}{2007}), \eprint{0705.3980}.

\bibitem[{\citenamefont{Hanson et~al.}(2013)}]{SPTpol:2013omd}
\bibinfo{author}{\bibfnamefont{D.}~\bibnamefont{Hanson}} \bibnamefont{et~al.}
  (\bibinfo{collaboration}{SPTpol}), \bibinfo{journal}{Phys. Rev. Lett.}
  \textbf{\bibinfo{volume}{111}}, \bibinfo{pages}{141301}
  (\bibinfo{year}{2013}), \eprint{1307.5830}.

\bibitem[{\citenamefont{Hand et~al.}(2012)}]{Hand:2012ui}
\bibinfo{author}{\bibfnamefont{N.}~\bibnamefont{Hand}} \bibnamefont{et~al.},
  \bibinfo{journal}{Phys. Rev. Lett.} \textbf{\bibinfo{volume}{109}},
  \bibinfo{pages}{041101} (\bibinfo{year}{2012}), \eprint{1203.4219}.

\bibitem[{\citenamefont{{Frieman} et~al.}(2008)\citenamefont{{Frieman},
  {Turner}, and {Huterer}}}]{2008ARA&A..46..385F}
\bibinfo{author}{\bibfnamefont{J.~A.} \bibnamefont{{Frieman}}},
  \bibinfo{author}{\bibfnamefont{M.~S.} \bibnamefont{{Turner}}},
  \bibnamefont{and}
  \bibinfo{author}{\bibfnamefont{D.}~\bibnamefont{{Huterer}}},
  \bibinfo{journal}{\araa} \textbf{\bibinfo{volume}{46}}, \bibinfo{pages}{385}
  (\bibinfo{year}{2008}), \eprint{0803.0982}.

\bibitem[{\citenamefont{{Weinberg} et~al.}(2013)\citenamefont{{Weinberg},
  {Mortonson}, {Eisenstein}, {Hirata}, {Riess}, and {Rozo}}}]{Weinberg2013}
\bibinfo{author}{\bibfnamefont{D.~H.} \bibnamefont{{Weinberg}}},
  \bibinfo{author}{\bibfnamefont{M.~J.} \bibnamefont{{Mortonson}}},
  \bibinfo{author}{\bibfnamefont{D.~J.} \bibnamefont{{Eisenstein}}},
  \bibinfo{author}{\bibfnamefont{C.}~\bibnamefont{{Hirata}}},
  \bibinfo{author}{\bibfnamefont{A.~G.} \bibnamefont{{Riess}}},
  \bibnamefont{and} \bibinfo{author}{\bibfnamefont{E.}~\bibnamefont{{Rozo}}},
  \bibinfo{journal}{\physrep} \textbf{\bibinfo{volume}{530}},
  \bibinfo{pages}{87} (\bibinfo{year}{2013}), \eprint{1201.2434}.

\bibitem[{\citenamefont{{Joyce} et~al.}(2016)\citenamefont{{Joyce},
  {Lombriser}, and {Schmidt}}}]{2016ARNPS..66...95J}
\bibinfo{author}{\bibfnamefont{A.}~\bibnamefont{{Joyce}}},
  \bibinfo{author}{\bibfnamefont{L.}~\bibnamefont{{Lombriser}}},
  \bibnamefont{and}
  \bibinfo{author}{\bibfnamefont{F.}~\bibnamefont{{Schmidt}}},
  \bibinfo{journal}{Annual Review of Nuclear and Particle Science}
  \textbf{\bibinfo{volume}{66}}, \bibinfo{pages}{95} (\bibinfo{year}{2016}),
  \eprint{1601.06133}.

\bibitem[{\citenamefont{{Desai}}(2018)}]{Desai18}
\bibinfo{author}{\bibfnamefont{S.}~\bibnamefont{{Desai}}},
  \bibinfo{journal}{Physics Letters B} \textbf{\bibinfo{volume}{778}},
  \bibinfo{pages}{325} (\bibinfo{year}{2018}), \eprint{1708.06502}.

\bibitem[{\citenamefont{King et~al.}(2012)\citenamefont{King, Webb, Murphy,
  Flambaum, Carswell, Bainbridge, Wilczynska, and Koch}}]{King:2012id}
\bibinfo{author}{\bibfnamefont{J.~A.} \bibnamefont{King}},
  \bibinfo{author}{\bibfnamefont{J.~K.} \bibnamefont{Webb}},
  \bibinfo{author}{\bibfnamefont{M.~T.} \bibnamefont{Murphy}},
  \bibinfo{author}{\bibfnamefont{V.~V.} \bibnamefont{Flambaum}},
  \bibinfo{author}{\bibfnamefont{R.~F.} \bibnamefont{Carswell}},
  \bibinfo{author}{\bibfnamefont{M.~B.} \bibnamefont{Bainbridge}},
  \bibinfo{author}{\bibfnamefont{M.~R.} \bibnamefont{Wilczynska}},
  \bibnamefont{and} \bibinfo{author}{\bibfnamefont{F.~E.} \bibnamefont{Koch}},
  \bibinfo{journal}{Mon. Not. Roy. Astron. Soc.}
  \textbf{\bibinfo{volume}{422}}, \bibinfo{pages}{3370} (\bibinfo{year}{2012}),
  \eprint{1202.4758}.

\bibitem[{\citenamefont{Leefer et~al.}(2013)\citenamefont{Leefer, Weber,
  Cing\"oz, Torgerson, and Budker}}]{Leefer:2013waa}
\bibinfo{author}{\bibfnamefont{N.}~\bibnamefont{Leefer}},
  \bibinfo{author}{\bibfnamefont{C.~T.~M.} \bibnamefont{Weber}},
  \bibinfo{author}{\bibfnamefont{A.}~\bibnamefont{Cing\"oz}},
  \bibinfo{author}{\bibfnamefont{J.~R.} \bibnamefont{Torgerson}},
  \bibnamefont{and} \bibinfo{author}{\bibfnamefont{D.}~\bibnamefont{Budker}},
  \bibinfo{journal}{Phys. Rev. Lett.} \textbf{\bibinfo{volume}{111}},
  \bibinfo{pages}{060801} (\bibinfo{year}{2013}), \eprint{1304.6940}.

\bibitem[{\citenamefont{van~de Bruck et~al.}(2015)\citenamefont{van~de Bruck,
  Mifsud, and Nunes}}]{vandeBruck:2015rma}
\bibinfo{author}{\bibfnamefont{C.}~\bibnamefont{van~de Bruck}},
  \bibinfo{author}{\bibfnamefont{J.}~\bibnamefont{Mifsud}}, \bibnamefont{and}
  \bibinfo{author}{\bibfnamefont{N.~J.} \bibnamefont{Nunes}},
  \bibinfo{journal}{JCAP} \textbf{\bibinfo{volume}{12}}, \bibinfo{pages}{018}
  (\bibinfo{year}{2015}), \eprint{1510.00200}.

\bibitem[{\citenamefont{Kotu\v{s} et~al.}(2017)\citenamefont{Kotu\v{s}, Murphy,
  and Carswell}}]{Kotus:2016xxb}
\bibinfo{author}{\bibfnamefont{S.~M.} \bibnamefont{Kotu\v{s}}},
  \bibinfo{author}{\bibfnamefont{M.~T.} \bibnamefont{Murphy}},
  \bibnamefont{and} \bibinfo{author}{\bibfnamefont{R.~F.}
  \bibnamefont{Carswell}}, \bibinfo{journal}{Mon. Not. Roy. Astron. Soc.}
  \textbf{\bibinfo{volume}{464}}, \bibinfo{pages}{3679} (\bibinfo{year}{2017}),
  \eprint{1609.03860}.

\bibitem[{\citenamefont{Gon\c{c}alves et~al.}(2020)\citenamefont{Gon\c{c}alves,
  Landau, Alcaniz, and Holanda}}]{Goncalves:2019xtc}
\bibinfo{author}{\bibfnamefont{R.~S.} \bibnamefont{Gon\c{c}alves}},
  \bibinfo{author}{\bibfnamefont{S.}~\bibnamefont{Landau}},
  \bibinfo{author}{\bibfnamefont{J.~S.} \bibnamefont{Alcaniz}},
  \bibnamefont{and} \bibinfo{author}{\bibfnamefont{R.~F.~L.}
  \bibnamefont{Holanda}}, \bibinfo{journal}{JCAP}
  \textbf{\bibinfo{volume}{06}}, \bibinfo{pages}{036} (\bibinfo{year}{2020}),
  \eprint{1907.02118}.

\bibitem[{\citenamefont{Galli}(2013)}]{Galli:2012bf}
\bibinfo{author}{\bibfnamefont{S.}~\bibnamefont{Galli}},
  \bibinfo{journal}{Phys. Rev. D} \textbf{\bibinfo{volume}{87}},
  \bibinfo{pages}{123516} (\bibinfo{year}{2013}), \eprint{1212.1075}.

\bibitem[{\citenamefont{Liu et~al.}(2021)\citenamefont{Liu, Liu, Zhang, Zhai,
  and Bora}}]{Liu:2021mfk}
\bibinfo{author}{\bibfnamefont{Z.-E.} \bibnamefont{Liu}},
  \bibinfo{author}{\bibfnamefont{W.-F.} \bibnamefont{Liu}},
  \bibinfo{author}{\bibfnamefont{T.-J.} \bibnamefont{Zhang}},
  \bibinfo{author}{\bibfnamefont{Z.-X.} \bibnamefont{Zhai}}, \bibnamefont{and}
  \bibinfo{author}{\bibfnamefont{K.}~\bibnamefont{Bora}}
  (\bibinfo{year}{2021}), \eprint{2109.00134}.

\bibitem[{\citenamefont{Bora and Desai}(2021)}]{Bora:2020sws}
\bibinfo{author}{\bibfnamefont{K.}~\bibnamefont{Bora}} \bibnamefont{and}
  \bibinfo{author}{\bibfnamefont{S.}~\bibnamefont{Desai}},
  \bibinfo{journal}{JCAP} \textbf{\bibinfo{volume}{02}}, \bibinfo{pages}{012}
  (\bibinfo{year}{2021}), \eprint{2008.10541}.

\bibitem[{\citenamefont{Cola\c{c}o et~al.}(2019)\citenamefont{Cola\c{c}o,
  Holanda, Silva, and Alcaniz}}]{Colaco:2019fvl}
\bibinfo{author}{\bibfnamefont{L.~R.} \bibnamefont{Cola\c{c}o}},
  \bibinfo{author}{\bibfnamefont{R.~F.~L.} \bibnamefont{Holanda}},
  \bibinfo{author}{\bibfnamefont{R.}~\bibnamefont{Silva}}, \bibnamefont{and}
  \bibinfo{author}{\bibfnamefont{J.~S.} \bibnamefont{Alcaniz}},
  \bibinfo{journal}{JCAP} \textbf{\bibinfo{volume}{03}}, \bibinfo{pages}{014}
  (\bibinfo{year}{2019}), \eprint{1901.10947}.

\bibitem[{\citenamefont{{Dirac}}(1937)}]{Dirac}
\bibinfo{author}{\bibfnamefont{P.~A.~M.} \bibnamefont{{Dirac}}},
  \bibinfo{journal}{\nat} \textbf{\bibinfo{volume}{139}}, \bibinfo{pages}{323}
  (\bibinfo{year}{1937}).

\bibitem[{\citenamefont{Jofre et~al.}(2006)\citenamefont{Jofre, Reisenegger,
  and Fernandez}}]{Jofre:2006ug}
\bibinfo{author}{\bibfnamefont{P.}~\bibnamefont{Jofre}},
  \bibinfo{author}{\bibfnamefont{A.}~\bibnamefont{Reisenegger}},
  \bibnamefont{and}
  \bibinfo{author}{\bibfnamefont{R.}~\bibnamefont{Fernandez}},
  \bibinfo{journal}{Phys. Rev. Lett.} \textbf{\bibinfo{volume}{97}},
  \bibinfo{pages}{131102} (\bibinfo{year}{2006}), \eprint{astro-ph/0606708}.

\bibitem[{\citenamefont{Verbiest et~al.}(2008)\citenamefont{Verbiest, Bailes,
  van Straten, Hobbs, Edwards, Manchester, Bhat, Sarkissian, Jacoby, and
  Kulkarni}}]{Verbiest:2008gy}
\bibinfo{author}{\bibfnamefont{J.~P.~W.} \bibnamefont{Verbiest}},
  \bibinfo{author}{\bibfnamefont{M.}~\bibnamefont{Bailes}},
  \bibinfo{author}{\bibfnamefont{W.}~\bibnamefont{van Straten}},
  \bibinfo{author}{\bibfnamefont{G.~B.} \bibnamefont{Hobbs}},
  \bibinfo{author}{\bibfnamefont{R.~T.} \bibnamefont{Edwards}},
  \bibinfo{author}{\bibfnamefont{R.~N.} \bibnamefont{Manchester}},
  \bibinfo{author}{\bibfnamefont{N.~D.~R.} \bibnamefont{Bhat}},
  \bibinfo{author}{\bibfnamefont{J.~M.} \bibnamefont{Sarkissian}},
  \bibinfo{author}{\bibfnamefont{B.~A.} \bibnamefont{Jacoby}},
  \bibnamefont{and} \bibinfo{author}{\bibfnamefont{S.~R.}
  \bibnamefont{Kulkarni}}, \bibinfo{journal}{Astrophys. J.}
  \textbf{\bibinfo{volume}{679}}, \bibinfo{pages}{675} (\bibinfo{year}{2008}),
  \eprint{0801.2589}.

\bibitem[{\citenamefont{Lazaridis et~al.}(2009)}]{Lazaridis:2009kq}
\bibinfo{author}{\bibfnamefont{K.}~\bibnamefont{Lazaridis}}
  \bibnamefont{et~al.}, \bibinfo{journal}{Mon. Not. R. Astron. Soc.}
  \textbf{\bibinfo{volume}{400}}, \bibinfo{pages}{805} (\bibinfo{year}{2009}),
  \eprint{0908.0285}.

\bibitem[{\citenamefont{Garcia-Berro et~al.}(2011)\citenamefont{Garcia-Berro,
  Loren-Aguilar, Torres, Althaus, and Isern}}]{Garcia-Berro:2011kvq}
\bibinfo{author}{\bibfnamefont{E.}~\bibnamefont{Garcia-Berro}},
  \bibinfo{author}{\bibfnamefont{P.}~\bibnamefont{Loren-Aguilar}},
  \bibinfo{author}{\bibfnamefont{S.}~\bibnamefont{Torres}},
  \bibinfo{author}{\bibfnamefont{L.~G.} \bibnamefont{Althaus}},
  \bibnamefont{and} \bibinfo{author}{\bibfnamefont{J.}~\bibnamefont{Isern}},
  \bibinfo{journal}{JCAP} \textbf{\bibinfo{volume}{05}}, \bibinfo{pages}{021}
  (\bibinfo{year}{2011}), \eprint{1105.1992}.

\bibitem[{\citenamefont{Ooba et~al.}(2016)\citenamefont{Ooba, Ichiki, Chiba,
  and Sugiyama}}]{Ooba:2016slp}
\bibinfo{author}{\bibfnamefont{J.}~\bibnamefont{Ooba}},
  \bibinfo{author}{\bibfnamefont{K.}~\bibnamefont{Ichiki}},
  \bibinfo{author}{\bibfnamefont{T.}~\bibnamefont{Chiba}}, \bibnamefont{and}
  \bibinfo{author}{\bibfnamefont{N.}~\bibnamefont{Sugiyama}},
  \bibinfo{journal}{Phys. Rev. D} \textbf{\bibinfo{volume}{93}},
  \bibinfo{pages}{122002} (\bibinfo{year}{2016}), \eprint{1602.00809}.

\bibitem[{\citenamefont{Zhao et~al.}(2018)\citenamefont{Zhao, Wright, and
  Li}}]{Zhao:2018gwk}
\bibinfo{author}{\bibfnamefont{W.}~\bibnamefont{Zhao}},
  \bibinfo{author}{\bibfnamefont{B.~S.} \bibnamefont{Wright}},
  \bibnamefont{and} \bibinfo{author}{\bibfnamefont{B.}~\bibnamefont{Li}},
  \bibinfo{journal}{JCAP} \textbf{\bibinfo{volume}{10}}, \bibinfo{pages}{052}
  (\bibinfo{year}{2018}), \eprint{1804.03066}.

\bibitem[{\citenamefont{Vijaykumar et~al.}(2021)\citenamefont{Vijaykumar,
  Kapadia, and Ajith}}]{Vijaykumar:2020nzc}
\bibinfo{author}{\bibfnamefont{A.}~\bibnamefont{Vijaykumar}},
  \bibinfo{author}{\bibfnamefont{S.~J.} \bibnamefont{Kapadia}},
  \bibnamefont{and} \bibinfo{author}{\bibfnamefont{P.}~\bibnamefont{Ajith}},
  \bibinfo{journal}{Phys. Rev. Lett.} \textbf{\bibinfo{volume}{126}},
  \bibinfo{pages}{141104} (\bibinfo{year}{2021}), \eprint{2003.12832}.

\bibitem[{\citenamefont{Zhu and Ma}(2021)}]{Zhu:2021pml}
\bibinfo{author}{\bibfnamefont{J.}~\bibnamefont{Zhu}} \bibnamefont{and}
  \bibinfo{author}{\bibfnamefont{B.-Q.} \bibnamefont{Ma}},
  \bibinfo{journal}{Phys. Lett. B} \textbf{\bibinfo{volume}{820}},
  \bibinfo{pages}{136518} (\bibinfo{year}{2021}), \eprint{2108.05804}.

\bibitem[{\citenamefont{Liu and Ma}(2018)}]{Liu:2018qrg}
\bibinfo{author}{\bibfnamefont{Y.}~\bibnamefont{Liu}} \bibnamefont{and}
  \bibinfo{author}{\bibfnamefont{B.-Q.} \bibnamefont{Ma}},
  \bibinfo{journal}{Eur. Phys. J. C} \textbf{\bibinfo{volume}{78}},
  \bibinfo{pages}{825} (\bibinfo{year}{2018}), \eprint{1810.00636}.

\bibitem[{\citenamefont{Xu and Ma}(2016{\natexlab{a}})}]{Xu:2016zsa}
\bibinfo{author}{\bibfnamefont{H.}~\bibnamefont{Xu}} \bibnamefont{and}
  \bibinfo{author}{\bibfnamefont{B.-Q.} \bibnamefont{Ma}},
  \bibinfo{journal}{Phys. Lett. B} \textbf{\bibinfo{volume}{760}},
  \bibinfo{pages}{602} (\bibinfo{year}{2016}{\natexlab{a}}),
  \eprint{1607.08043}.

\bibitem[{\citenamefont{Xu and Ma}(2016{\natexlab{b}})}]{Xu:2016zxi}
\bibinfo{author}{\bibfnamefont{H.}~\bibnamefont{Xu}} \bibnamefont{and}
  \bibinfo{author}{\bibfnamefont{B.-Q.} \bibnamefont{Ma}},
  \bibinfo{journal}{Astropart. Phys.} \textbf{\bibinfo{volume}{82}},
  \bibinfo{pages}{72} (\bibinfo{year}{2016}{\natexlab{b}}),
  \eprint{1607.03203}.

\bibitem[{\citenamefont{Cruz and Faria}(2012)}]{Cruz:2012bwp}
\bibinfo{author}{\bibfnamefont{C.~N.} \bibnamefont{Cruz}} \bibnamefont{and}
  \bibinfo{author}{\bibfnamefont{A.~C. A.~d.} \bibnamefont{Faria}},
  \bibinfo{journal}{Phys. Rev. D} \textbf{\bibinfo{volume}{86}},
  \bibinfo{pages}{027703} (\bibinfo{year}{2012}), \eprint{1205.2298}.

\bibitem[{\citenamefont{Ackermann et~al.}(2009)}]{FermiGBMLAT:2009nfe}
\bibinfo{author}{\bibfnamefont{M.}~\bibnamefont{Ackermann}}
  \bibnamefont{et~al.} (\bibinfo{collaboration}{Fermi GBM/LAT}),
  \bibinfo{journal}{Nature} \textbf{\bibinfo{volume}{462}},
  \bibinfo{pages}{331} (\bibinfo{year}{2009}), \eprint{0908.1832}.

\bibitem[{\citenamefont{{Liu} et~al.}(2021)\citenamefont{{Liu}, {Cao},
  {Biesiada}, {Liu}, {Lian}, and {Zhang}}}]{Liu:2021eit}
\bibinfo{author}{\bibfnamefont{T.}~\bibnamefont{{Liu}}},
  \bibinfo{author}{\bibfnamefont{S.}~\bibnamefont{{Cao}}},
  \bibinfo{author}{\bibfnamefont{M.}~\bibnamefont{{Biesiada}}},
  \bibinfo{author}{\bibfnamefont{Y.}~\bibnamefont{{Liu}}},
  \bibinfo{author}{\bibfnamefont{Y.}~\bibnamefont{{Lian}}}, \bibnamefont{and}
  \bibinfo{author}{\bibfnamefont{Y.}~\bibnamefont{{Zhang}}},
  \bibinfo{journal}{\mnras} \textbf{\bibinfo{volume}{506}},
  \bibinfo{pages}{2181} (\bibinfo{year}{2021}), \eprint{2106.15145}.

\bibitem[{\citenamefont{Cao et~al.}(2018)\citenamefont{Cao, Qi, Biesiada,
  Zheng, Xu, and Zhu}}]{Cao:2018rzc}
\bibinfo{author}{\bibfnamefont{S.}~\bibnamefont{Cao}},
  \bibinfo{author}{\bibfnamefont{J.}~\bibnamefont{Qi}},
  \bibinfo{author}{\bibfnamefont{M.}~\bibnamefont{Biesiada}},
  \bibinfo{author}{\bibfnamefont{X.}~\bibnamefont{Zheng}},
  \bibinfo{author}{\bibfnamefont{T.}~\bibnamefont{Xu}}, \bibnamefont{and}
  \bibinfo{author}{\bibfnamefont{Z.-H.} \bibnamefont{Zhu}},
  \bibinfo{journal}{Astrophys. J.} \textbf{\bibinfo{volume}{867}},
  \bibinfo{pages}{50} (\bibinfo{year}{2018}), \eprint{1810.01287}.

\bibitem[{\citenamefont{Cao et~al.}(2017)\citenamefont{Cao, Biesiada, Jackson,
  Zheng, Zhao, and Zhu}}]{Cao:2016dgw}
\bibinfo{author}{\bibfnamefont{S.}~\bibnamefont{Cao}},
  \bibinfo{author}{\bibfnamefont{M.}~\bibnamefont{Biesiada}},
  \bibinfo{author}{\bibfnamefont{J.}~\bibnamefont{Jackson}},
  \bibinfo{author}{\bibfnamefont{X.}~\bibnamefont{Zheng}},
  \bibinfo{author}{\bibfnamefont{Y.}~\bibnamefont{Zhao}}, \bibnamefont{and}
  \bibinfo{author}{\bibfnamefont{Z.-H.} \bibnamefont{Zhu}},
  \bibinfo{journal}{JCAP} \textbf{\bibinfo{volume}{02}}, \bibinfo{pages}{012}
  (\bibinfo{year}{2017}), \eprint{1609.08748}.

\bibitem[{\citenamefont{{Agrawal} et~al.}(2021)\citenamefont{{Agrawal},
  {Singirikonda}, and {Desai}}}]{Rajdeep}
\bibinfo{author}{\bibfnamefont{R.}~\bibnamefont{{Agrawal}}},
  \bibinfo{author}{\bibfnamefont{H.}~\bibnamefont{{Singirikonda}}},
  \bibnamefont{and} \bibinfo{author}{\bibfnamefont{S.}~\bibnamefont{{Desai}}},
  \bibinfo{journal}{\jcap} \textbf{\bibinfo{volume}{2021}}, \bibinfo{eid}{029}
  (\bibinfo{year}{2021}), \eprint{2102.11248}.

\bibitem[{\citenamefont{{Moffat}}(1993)}]{Moffat93}
\bibinfo{author}{\bibfnamefont{J.~W.} \bibnamefont{{Moffat}}},
  \bibinfo{journal}{International Journal of Modern Physics D}
  \textbf{\bibinfo{volume}{2}}, \bibinfo{pages}{351} (\bibinfo{year}{1993}),
  \eprint{gr-qc/9211020}.

\bibitem[{\citenamefont{Albrecht and Magueijo}(1999)}]{Albrecht:1998ir}
\bibinfo{author}{\bibfnamefont{A.}~\bibnamefont{Albrecht}} \bibnamefont{and}
  \bibinfo{author}{\bibfnamefont{J.}~\bibnamefont{Magueijo}},
  \bibinfo{journal}{Phys. Rev. D} \textbf{\bibinfo{volume}{59}},
  \bibinfo{pages}{043516} (\bibinfo{year}{1999}), \eprint{astro-ph/9811018}.

\bibitem[{\citenamefont{Magueijo}(2003)}]{Magueijo:2003gj}
\bibinfo{author}{\bibfnamefont{J.}~\bibnamefont{Magueijo}},
  \bibinfo{journal}{Rept. Prog. Phys.} \textbf{\bibinfo{volume}{66}},
  \bibinfo{pages}{2025} (\bibinfo{year}{2003}), \eprint{astro-ph/0305457}.

\bibitem[{\citenamefont{Ellis and Uzan}(2005)}]{Ellis:2003pw}
\bibinfo{author}{\bibfnamefont{G.~F.~R.} \bibnamefont{Ellis}} \bibnamefont{and}
  \bibinfo{author}{\bibfnamefont{J.-P.} \bibnamefont{Uzan}},
  \bibinfo{journal}{Am. J. Phys.} \textbf{\bibinfo{volume}{73}},
  \bibinfo{pages}{240} (\bibinfo{year}{2005}), \eprint{gr-qc/0305099}.

\bibitem[{\citenamefont{{Ellis}}(2007)}]{Ellis07}
\bibinfo{author}{\bibfnamefont{G.~F.~R.} \bibnamefont{{Ellis}}},
  \bibinfo{journal}{General Relativity and Gravitation}
  \textbf{\bibinfo{volume}{39}}, \bibinfo{pages}{511} (\bibinfo{year}{2007}),
  \eprint{astro-ph/0703751}.

\bibitem[{\citenamefont{Duff et~al.}(2002)\citenamefont{Duff, Okun, and
  Veneziano}}]{Duff:2001ba}
\bibinfo{author}{\bibfnamefont{M.~J.} \bibnamefont{Duff}},
  \bibinfo{author}{\bibfnamefont{L.~B.} \bibnamefont{Okun}}, \bibnamefont{and}
  \bibinfo{author}{\bibfnamefont{G.}~\bibnamefont{Veneziano}},
  \bibinfo{journal}{JHEP} \textbf{\bibinfo{volume}{03}}, \bibinfo{pages}{023}
  (\bibinfo{year}{2002}), \eprint{physics/0110060}.

\bibitem[{\citenamefont{Uzan}(2003)}]{Uzan:2002vq}
\bibinfo{author}{\bibfnamefont{J.-P.} \bibnamefont{Uzan}},
  \bibinfo{journal}{Rev. Mod. Phys.} \textbf{\bibinfo{volume}{75}},
  \bibinfo{pages}{403} (\bibinfo{year}{2003}), \eprint{hep-ph/0205340}.

\bibitem[{\citenamefont{{Magueijo} and {Moffat}}(2008)}]{Moffat08}
\bibinfo{author}{\bibfnamefont{J.}~\bibnamefont{{Magueijo}}} \bibnamefont{and}
  \bibinfo{author}{\bibfnamefont{J.~W.} \bibnamefont{{Moffat}}},
  \bibinfo{journal}{General Relativity and Gravitation}
  \textbf{\bibinfo{volume}{40}}, \bibinfo{pages}{1797} (\bibinfo{year}{2008}),
  \eprint{0705.4507}.

\bibitem[{\citenamefont{Salzano et~al.}(2015)\citenamefont{Salzano, Dabrowski,
  and Lazkoz}}]{Salzano:2014lra}
\bibinfo{author}{\bibfnamefont{V.}~\bibnamefont{Salzano}},
  \bibinfo{author}{\bibfnamefont{M.~P.} \bibnamefont{Dabrowski}},
  \bibnamefont{and} \bibinfo{author}{\bibfnamefont{R.}~\bibnamefont{Lazkoz}},
  \bibinfo{journal}{Phys. Rev. Lett.} \textbf{\bibinfo{volume}{114}},
  \bibinfo{pages}{101304} (\bibinfo{year}{2015}), \eprint{1412.5653}.

\bibitem[{\citenamefont{Salzano}(2017)}]{Salzano:2016hce}
\bibinfo{author}{\bibfnamefont{V.}~\bibnamefont{Salzano}},
  \bibinfo{journal}{Phys. Rev. D} \textbf{\bibinfo{volume}{95}},
  \bibinfo{pages}{084035} (\bibinfo{year}{2017}), \eprint{1604.03398}.

\bibitem[{\citenamefont{Qi et~al.}(2014)\citenamefont{Qi, Zhang, and
  Liu}}]{Qi:2014zja}
\bibinfo{author}{\bibfnamefont{J.-Z.} \bibnamefont{Qi}},
  \bibinfo{author}{\bibfnamefont{M.-J.} \bibnamefont{Zhang}}, \bibnamefont{and}
  \bibinfo{author}{\bibfnamefont{W.-B.} \bibnamefont{Liu}},
  \bibinfo{journal}{Phys. Rev. D} \textbf{\bibinfo{volume}{90}},
  \bibinfo{pages}{063526} (\bibinfo{year}{2014}), \eprint{1407.1265}.

\bibitem[{\citenamefont{Wang et~al.}(2019)\citenamefont{Wang, Zhang, Zheng,
  Wang, and Zhao}}]{Wang:2019tdn}
\bibinfo{author}{\bibfnamefont{D.}~\bibnamefont{Wang}},
  \bibinfo{author}{\bibfnamefont{H.}~\bibnamefont{Zhang}},
  \bibinfo{author}{\bibfnamefont{J.}~\bibnamefont{Zheng}},
  \bibinfo{author}{\bibfnamefont{Y.}~\bibnamefont{Wang}}, \bibnamefont{and}
  \bibinfo{author}{\bibfnamefont{G.-B.} \bibnamefont{Zhao}}
  (\bibinfo{year}{2019}), \eprint{1904.04041}.

\bibitem[{\citenamefont{{Bartlett} et~al.}(2021)\citenamefont{{Bartlett},
  {Desmond}, {Ferreira}, and {Jasche}}}]{Bartlett}
\bibinfo{author}{\bibfnamefont{D.~J.} \bibnamefont{{Bartlett}}},
  \bibinfo{author}{\bibfnamefont{H.}~\bibnamefont{{Desmond}}},
  \bibinfo{author}{\bibfnamefont{P.~G.} \bibnamefont{{Ferreira}}},
  \bibnamefont{and} \bibinfo{author}{\bibfnamefont{J.}~\bibnamefont{{Jasche}}},
  \bibinfo{journal}{arXiv e-prints} \bibinfo{eid}{arXiv:2109.07850}
  (\bibinfo{year}{2021}), \eprint{2109.07850}.

\bibitem[{\citenamefont{Mantz et~al.}(2014)\citenamefont{Mantz, Allen, Morris,
  Rapetti, Applegate, Kelly, von~der Linden, and Schmidt}}]{Mantz:2014xba}
\bibinfo{author}{\bibfnamefont{A.~B.} \bibnamefont{Mantz}},
  \bibinfo{author}{\bibfnamefont{S.~W.} \bibnamefont{Allen}},
  \bibinfo{author}{\bibfnamefont{R.~G.} \bibnamefont{Morris}},
  \bibinfo{author}{\bibfnamefont{D.~A.} \bibnamefont{Rapetti}},
  \bibinfo{author}{\bibfnamefont{D.~E.} \bibnamefont{Applegate}},
  \bibinfo{author}{\bibfnamefont{P.~L.} \bibnamefont{Kelly}},
  \bibinfo{author}{\bibfnamefont{A.}~\bibnamefont{von~der Linden}},
  \bibnamefont{and} \bibinfo{author}{\bibfnamefont{R.~W.}
  \bibnamefont{Schmidt}}, \bibinfo{journal}{Mon. Not. Roy. Astron. Soc.}
  \textbf{\bibinfo{volume}{440}}, \bibinfo{pages}{2077} (\bibinfo{year}{2014}),
  \eprint{1402.6212}.

\bibitem[{\citenamefont{{Li} et~al.}(2021)\citenamefont{{Li}, {Du}, {Zhou},
  {Zhang}, and {Xu}}}]{li19}
\bibinfo{author}{\bibfnamefont{E.-K.} \bibnamefont{{Li}}},
  \bibinfo{author}{\bibfnamefont{M.}~\bibnamefont{{Du}}},
  \bibinfo{author}{\bibfnamefont{Z.-H.} \bibnamefont{{Zhou}}},
  \bibinfo{author}{\bibfnamefont{H.}~\bibnamefont{{Zhang}}}, \bibnamefont{and}
  \bibinfo{author}{\bibfnamefont{L.}~\bibnamefont{{Xu}}},
  \bibinfo{journal}{\mnras} \textbf{\bibinfo{volume}{501}},
  \bibinfo{pages}{4452} (\bibinfo{year}{2021}), \eprint{1911.12076}.

\bibitem[{\citenamefont{{Scolnic} et~al.}(2018)\citenamefont{{Scolnic},
  {Jones}, {Rest}, {Pan}, {Chornock}, {Foley}, {Huber}, {Kessler}, {Narayan},
  {Riess} et~al.}}]{pantheon}
\bibinfo{author}{\bibfnamefont{D.~M.} \bibnamefont{{Scolnic}}},
  \bibinfo{author}{\bibfnamefont{D.~O.} \bibnamefont{{Jones}}},
  \bibinfo{author}{\bibfnamefont{A.}~\bibnamefont{{Rest}}},
  \bibinfo{author}{\bibfnamefont{Y.~C.} \bibnamefont{{Pan}}},
  \bibinfo{author}{\bibfnamefont{R.}~\bibnamefont{{Chornock}}},
  \bibinfo{author}{\bibfnamefont{R.~J.} \bibnamefont{{Foley}}},
  \bibinfo{author}{\bibfnamefont{M.~E.} \bibnamefont{{Huber}}},
  \bibinfo{author}{\bibfnamefont{R.}~\bibnamefont{{Kessler}}},
  \bibinfo{author}{\bibfnamefont{G.}~\bibnamefont{{Narayan}}},
  \bibinfo{author}{\bibfnamefont{A.~G.} \bibnamefont{{Riess}}},
  \bibnamefont{et~al.}, \bibinfo{journal}{\apj} \textbf{\bibinfo{volume}{859}},
  \bibinfo{eid}{101} (\bibinfo{year}{2018}), \eprint{1710.00845}.

\bibitem[{\citenamefont{{Riess} et~al.}(2021)\citenamefont{{Riess},
  {Casertano}, {Yuan}, {Bowers}, {Macri}, {Zinn}, and
  {Scolnic}}}]{2021ApJ...908L...6R}
\bibinfo{author}{\bibfnamefont{A.~G.} \bibnamefont{{Riess}}},
  \bibinfo{author}{\bibfnamefont{S.}~\bibnamefont{{Casertano}}},
  \bibinfo{author}{\bibfnamefont{W.}~\bibnamefont{{Yuan}}},
  \bibinfo{author}{\bibfnamefont{J.~B.} \bibnamefont{{Bowers}}},
  \bibinfo{author}{\bibfnamefont{L.}~\bibnamefont{{Macri}}},
  \bibinfo{author}{\bibfnamefont{J.~C.} \bibnamefont{{Zinn}}},
  \bibnamefont{and}
  \bibinfo{author}{\bibfnamefont{D.}~\bibnamefont{{Scolnic}}},
  \bibinfo{journal}{\apjl} \textbf{\bibinfo{volume}{908}}, \bibinfo{eid}{L6}
  (\bibinfo{year}{2021}), \eprint{2012.08534}.

\bibitem[{\citenamefont{{Sarazin}}(1988)}]{sarazin}
\bibinfo{author}{\bibfnamefont{C.~L.} \bibnamefont{{Sarazin}}},
  \emph{\bibinfo{title}{{X-ray emission from clusters of galaxies}}}
  (\bibinfo{year}{1988}).

\bibitem[{\citenamefont{{Allen} et~al.}(2011)\citenamefont{{Allen}, {Evrard},
  and {Mantz}}}]{Allen2011}
\bibinfo{author}{\bibfnamefont{S.~W.} \bibnamefont{{Allen}}},
  \bibinfo{author}{\bibfnamefont{A.~E.} \bibnamefont{{Evrard}}},
  \bibnamefont{and} \bibinfo{author}{\bibfnamefont{A.~B.}
  \bibnamefont{{Mantz}}}, \bibinfo{journal}{\araa}
  \textbf{\bibinfo{volume}{49}}, \bibinfo{pages}{409} (\bibinfo{year}{2011}),
  \eprint{1103.4829}.

\bibitem[{\citenamefont{Allen et~al.}(2007)\citenamefont{Allen, Rapetti,
  Schmidt, Ebeling, Morris, and Fabian}}]{Allen2007}
\bibinfo{author}{\bibfnamefont{S.~W.} \bibnamefont{Allen}},
  \bibinfo{author}{\bibfnamefont{D.~A.} \bibnamefont{Rapetti}},
  \bibinfo{author}{\bibfnamefont{R.~W.} \bibnamefont{Schmidt}},
  \bibinfo{author}{\bibfnamefont{H.}~\bibnamefont{Ebeling}},
  \bibinfo{author}{\bibfnamefont{R.~G.} \bibnamefont{Morris}},
  \bibnamefont{and} \bibinfo{author}{\bibfnamefont{A.~C.}
  \bibnamefont{Fabian}}, \bibinfo{journal}{Monthly Notices of the Royal
  Astronomical Society} \textbf{\bibinfo{volume}{383}}, \bibinfo{pages}{879}
  (\bibinfo{year}{2007}).

\bibitem[{\citenamefont{{Ettori} et~al.}(2009)\citenamefont{{Ettori},
  {Morandi}, {Tozzi}, {Balestra}, {Borgani}, {Rosati}, {Lovisari}, and
  {Terenziani}}}]{Ettori2009}
\bibinfo{author}{\bibfnamefont{S.}~\bibnamefont{{Ettori}}},
  \bibinfo{author}{\bibfnamefont{A.}~\bibnamefont{{Morandi}}},
  \bibinfo{author}{\bibfnamefont{P.}~\bibnamefont{{Tozzi}}},
  \bibinfo{author}{\bibfnamefont{I.}~\bibnamefont{{Balestra}}},
  \bibinfo{author}{\bibfnamefont{S.}~\bibnamefont{{Borgani}}},
  \bibinfo{author}{\bibfnamefont{P.}~\bibnamefont{{Rosati}}},
  \bibinfo{author}{\bibfnamefont{L.}~\bibnamefont{{Lovisari}}},
  \bibnamefont{and}
  \bibinfo{author}{\bibfnamefont{F.}~\bibnamefont{{Terenziani}}},
  \bibinfo{journal}{\aap} \textbf{\bibinfo{volume}{501}}, \bibinfo{pages}{61}
  (\bibinfo{year}{2009}), \eprint{0904.2740}.

\bibitem[{\citenamefont{Holanda et~al.}(2020)\citenamefont{Holanda, da~Silva,
  and Pereira}}]{Holanda2020}
\bibinfo{author}{\bibfnamefont{R.}~\bibnamefont{Holanda}},
  \bibinfo{author}{\bibfnamefont{G.~P.} \bibnamefont{da~Silva}},
  \bibnamefont{and} \bibinfo{author}{\bibfnamefont{S.}~\bibnamefont{Pereira}},
  \bibinfo{journal}{Journal of Cosmology and Astroparticle Physics}
  \textbf{\bibinfo{volume}{2020}}, \bibinfo{pages}{053} (\bibinfo{year}{2020}),
  \urlprefix\url{https://doi.org/10.1088/1475-7516/2020/09/053}.

\bibitem[{\citenamefont{{Battaglia} et~al.}(2013)\citenamefont{{Battaglia},
  {Bond}, {Pfrommer}, and {Sievers}}}]{2013ApJ...777..123B}
\bibinfo{author}{\bibfnamefont{N.}~\bibnamefont{{Battaglia}}},
  \bibinfo{author}{\bibfnamefont{J.~R.} \bibnamefont{{Bond}}},
  \bibinfo{author}{\bibfnamefont{C.}~\bibnamefont{{Pfrommer}}},
  \bibnamefont{and} \bibinfo{author}{\bibfnamefont{J.~L.}
  \bibnamefont{{Sievers}}}, \bibinfo{journal}{\apj}
  \textbf{\bibinfo{volume}{777}}, \bibinfo{eid}{123} (\bibinfo{year}{2013}),
  \eprint{1209.4082}.

\bibitem[{\citenamefont{{Planelles} et~al.}(2013)\citenamefont{{Planelles},
  {Borgani}, {Dolag}, {Ettori}, {Fabjan}, {Murante}, and
  {Tornatore}}}]{2013MNRAS.431.1487P}
\bibinfo{author}{\bibfnamefont{S.}~\bibnamefont{{Planelles}}},
  \bibinfo{author}{\bibfnamefont{S.}~\bibnamefont{{Borgani}}},
  \bibinfo{author}{\bibfnamefont{K.}~\bibnamefont{{Dolag}}},
  \bibinfo{author}{\bibfnamefont{S.}~\bibnamefont{{Ettori}}},
  \bibinfo{author}{\bibfnamefont{D.}~\bibnamefont{{Fabjan}}},
  \bibinfo{author}{\bibfnamefont{G.}~\bibnamefont{{Murante}}},
  \bibnamefont{and}
  \bibinfo{author}{\bibfnamefont{L.}~\bibnamefont{{Tornatore}}},
  \bibinfo{journal}{\mnras} \textbf{\bibinfo{volume}{431}},
  \bibinfo{pages}{1487} (\bibinfo{year}{2013}), \eprint{1209.5058}.

\bibitem[{\citenamefont{{Bora} and {Desai}}(2021{\natexlab{a}})}]{BoraDesai21}
\bibinfo{author}{\bibfnamefont{K.}~\bibnamefont{{Bora}}} \bibnamefont{and}
  \bibinfo{author}{\bibfnamefont{S.}~\bibnamefont{{Desai}}},
  \bibinfo{journal}{European Physical Journal C} \textbf{\bibinfo{volume}{81}},
  \bibinfo{eid}{296} (\bibinfo{year}{2021}{\natexlab{a}}), \eprint{2103.12695}.

\bibitem[{\citenamefont{{Zheng} et~al.}(2019)\citenamefont{{Zheng}, {Qi},
  {Cao}, {Liu}, {Biesiada}, {Miernik}, and {Zhu}}}]{Zheng19}
\bibinfo{author}{\bibfnamefont{X.}~\bibnamefont{{Zheng}}},
  \bibinfo{author}{\bibfnamefont{J.-Z.} \bibnamefont{{Qi}}},
  \bibinfo{author}{\bibfnamefont{S.}~\bibnamefont{{Cao}}},
  \bibinfo{author}{\bibfnamefont{T.}~\bibnamefont{{Liu}}},
  \bibinfo{author}{\bibfnamefont{M.}~\bibnamefont{{Biesiada}}},
  \bibinfo{author}{\bibfnamefont{S.}~\bibnamefont{{Miernik}}},
  \bibnamefont{and} \bibinfo{author}{\bibfnamefont{Z.-H.} \bibnamefont{{Zhu}}},
  \bibinfo{journal}{European Physical Journal C} \textbf{\bibinfo{volume}{79}},
  \bibinfo{eid}{637} (\bibinfo{year}{2019}), \eprint{1907.06509}.

\bibitem[{\citenamefont{{Holanda} et~al.}(2017)\citenamefont{{Holanda},
  {Busti}, {Gonzalez}, {Andrade-Santos}, and {Alcaniz}}}]{2017JCAP...12..016H}
\bibinfo{author}{\bibfnamefont{R.~F.~L.} \bibnamefont{{Holanda}}},
  \bibinfo{author}{\bibfnamefont{V.~C.} \bibnamefont{{Busti}}},
  \bibinfo{author}{\bibfnamefont{J.~E.} \bibnamefont{{Gonzalez}}},
  \bibinfo{author}{\bibfnamefont{F.}~\bibnamefont{{Andrade-Santos}}},
  \bibnamefont{and} \bibinfo{author}{\bibfnamefont{J.~S.}
  \bibnamefont{{Alcaniz}}}, \bibinfo{journal}{\jcap}
  \textbf{\bibinfo{volume}{2017}}, \bibinfo{eid}{016} (\bibinfo{year}{2017}),
  \eprint{1706.07321}.

\bibitem[{\citenamefont{{Holanda} et~al.}(2021)\citenamefont{{Holanda}, {Bora},
  and {Desai}}}]{Holanda21}
\bibinfo{author}{\bibfnamefont{R.~F.~L.} \bibnamefont{{Holanda}}},
  \bibinfo{author}{\bibfnamefont{K.}~\bibnamefont{{Bora}}}, \bibnamefont{and}
  \bibinfo{author}{\bibfnamefont{S.}~\bibnamefont{{Desai}}},
  \bibinfo{journal}{arXiv e-prints} \bibinfo{eid}{arXiv:2105.10988}
  (\bibinfo{year}{2021}), \eprint{2105.10988}.

\bibitem[{\citenamefont{{Mendon{\c{c}}a}
  et~al.}(2021)\citenamefont{{Mendon{\c{c}}a}, {Bora}, {Holanda}, and
  {Desai}}}]{Mendonca21}
\bibinfo{author}{\bibfnamefont{I.~E.~C.~R.} \bibnamefont{{Mendon{\c{c}}a}}},
  \bibinfo{author}{\bibfnamefont{K.}~\bibnamefont{{Bora}}},
  \bibinfo{author}{\bibfnamefont{R.~F.~L.} \bibnamefont{{Holanda}}},
  \bibnamefont{and} \bibinfo{author}{\bibfnamefont{S.}~\bibnamefont{{Desai}}},
  \bibinfo{journal}{arXiv e-prints} \bibinfo{eid}{arXiv:2107.14169}
  (\bibinfo{year}{2021}), \eprint{2107.14169}.

\bibitem[{\citenamefont{{Jimenez} and {Loeb}}(2002)}]{jimenez}
\bibinfo{author}{\bibfnamefont{R.}~\bibnamefont{{Jimenez}}} \bibnamefont{and}
  \bibinfo{author}{\bibfnamefont{A.}~\bibnamefont{{Loeb}}},
  \bibinfo{journal}{\apj} \textbf{\bibinfo{volume}{573}}, \bibinfo{pages}{37}
  (\bibinfo{year}{2002}), \eprint{astro-ph/0106145}.

\bibitem[{\citenamefont{{Singirikonda} and {Desai}}(2020)}]{Haveesh}
\bibinfo{author}{\bibfnamefont{H.}~\bibnamefont{{Singirikonda}}}
  \bibnamefont{and} \bibinfo{author}{\bibfnamefont{S.}~\bibnamefont{{Desai}}},
  \bibinfo{journal}{European Physical Journal C} \textbf{\bibinfo{volume}{80}},
  \bibinfo{eid}{694} (\bibinfo{year}{2020}), \eprint{2003.00494}.

\bibitem[{\citenamefont{{Vagnozzi} et~al.}(2021)\citenamefont{{Vagnozzi},
  {Loeb}, and {Moresco}}}]{Vagnozzi}
\bibinfo{author}{\bibfnamefont{S.}~\bibnamefont{{Vagnozzi}}},
  \bibinfo{author}{\bibfnamefont{A.}~\bibnamefont{{Loeb}}}, \bibnamefont{and}
  \bibinfo{author}{\bibfnamefont{M.}~\bibnamefont{{Moresco}}},
  \bibinfo{journal}{\apj} \textbf{\bibinfo{volume}{908}}, \bibinfo{eid}{84}
  (\bibinfo{year}{2021}), \eprint{2011.11645}.

\bibitem[{\citenamefont{Seikel et~al.}(2012)\citenamefont{Seikel, Clarkson, and
  Smith}}]{Seikel}
\bibinfo{author}{\bibfnamefont{M.}~\bibnamefont{Seikel}},
  \bibinfo{author}{\bibfnamefont{C.}~\bibnamefont{Clarkson}}, \bibnamefont{and}
  \bibinfo{author}{\bibfnamefont{M.}~\bibnamefont{Smith}},
  \bibinfo{journal}{JCAP} \textbf{\bibinfo{volume}{1206}}, \bibinfo{pages}{036}
  (\bibinfo{year}{2012}), \eprint{1204.2832}.

\bibitem[{\citenamefont{{Bora} and
  {Desai}}(2021{\natexlab{b}})}]{BoraDesaiCDDR}
\bibinfo{author}{\bibfnamefont{K.}~\bibnamefont{{Bora}}} \bibnamefont{and}
  \bibinfo{author}{\bibfnamefont{S.}~\bibnamefont{{Desai}}},
  \bibinfo{journal}{\jcap} \textbf{\bibinfo{volume}{2021}}, \bibinfo{eid}{052}
  (\bibinfo{year}{2021}{\natexlab{b}}), \eprint{2104.00974}.

\bibitem[{\citenamefont{{Foreman-Mackey}
  et~al.}(2013)\citenamefont{{Foreman-Mackey}, {Hogg}, {Lang}, and
  {Goodman}}}]{emcee}
\bibinfo{author}{\bibfnamefont{D.}~\bibnamefont{{Foreman-Mackey}}},
  \bibinfo{author}{\bibfnamefont{D.~W.} \bibnamefont{{Hogg}}},
  \bibinfo{author}{\bibfnamefont{D.}~\bibnamefont{{Lang}}}, \bibnamefont{and}
  \bibinfo{author}{\bibfnamefont{J.}~\bibnamefont{{Goodman}}},
  \bibinfo{journal}{\pasp} \textbf{\bibinfo{volume}{125}}, \bibinfo{pages}{306}
  (\bibinfo{year}{2013}), \eprint{1202.3665}.

\bibitem[{\citenamefont{Foreman-Mackey}(2016)}]{corner}
\bibinfo{author}{\bibfnamefont{D.}~\bibnamefont{Foreman-Mackey}},
  \bibinfo{journal}{The Journal of Open Source Software}
  \textbf{\bibinfo{volume}{1}}, \bibinfo{pages}{24} (\bibinfo{year}{2016}),
  \urlprefix\url{https://doi.org/10.21105/joss.00024}.

\end{thebibliography}

\end{document}